\documentclass[pre,twocolumn,aps]{revtex4}
\usepackage{graphicx} 
\usepackage{bm}
\usepackage{amsmath}
\begin{document}

\title{Role of the first coordination shell in determining the equilibrium structure and dynamics of simple liquids}
\author{S{\o}ren Toxvaerd and Jeppe C. Dyre}
\affiliation{DNRF centre  ``Glass and Time,'' IMFUFA, Department of Sciences, Roskilde University, Postbox 260, DK-4000 Roskilde, Denmark}
\date{\today}
\begin{abstract}
The traditional view that the physical properties of a simple liquid are determined primarily by its repulsive forces was recently challenged by Berthier and Tarjus, who showed that in some cases ignoring the attractions leads to large errors in the dynamics [L. Berthier and G. Tarjus, Phys. Rev. Lett. {\bf 103}, 170601 (2009); J. Chem. Phys. \textbf{134} 214503 (2011)]. We present simulations of the standard Lennard-Jones liquid at several condensed-fluid state points, including a fairly low density state and a very high density state, as well as simulations of the Kob-Andersen binary Lennard-Jones mixture at several temperatures. By varying the range of the forces, results for the thermodynamics, dynamics, and structure  show that the determining factor for getting the correct statics and dynamics is not whether or not the attractive forces {\it per se} are included in the simulations. What matters is whether or not interactions are included from all particles within the first coordination shell (FCS) -- the attractive forces can thus be ignored, but only at extremely high densities. The recognition of the importance of a local shell in condensed fluids goes back to van der Waals; our results confirm this idea and thereby the basic picture of the old hole- and cell theories for simple condensed fluids.
\end{abstract}

\maketitle

\section{Introduction}

Ever since van der Waals \cite{vdW} in 1873 formulated his theory ``the continuity of the gaseous and liquid states'', the general understanding of simple fluids has been that repulsive and attractive forces give rise to separate modifications of the ideal-gas equation of state and to separate contributions to the free energy. The harsh, repulsive  forces reduce the available volume and determine the structure of the fluid and thus its configurational entropy; the weaker, longer-range attractive forces give rise to an energy effect that reduces pressure and energy compared to those of an ideal gas at the same temperature and density. With respect to the  effect of the longer-range attractive forces van der Waals imagined that he could (quoting from Ref. \cite{rowwidom}) "... define an element of volume in a liquid which is small compared ..." with "... the range of the intermolecular force, but large enough for it to contain sufficient molecules for us to assume that there is within it a uniform distribution of molecules of number density $\rho$". He  assumed that the main effect of the attractive forces originates from molecules within this local sphere and estimated this sphere to have a range of 3-6 \AA \cite {rowwidom}. Below we present evidence that this local volume, although it does not have a uniform distribution of molecules, may be identified with the first coordination shell (FCS).  

It is well known that the van der Waals equation of state reproduces the qualitative behavior of the fluid state \cite{Fisher,Widom,Widom1}. Van der Waals' idea about the role of the repulsive and attractive forces lies behind Zwanzig's high-temperature expansion \cite{Zwanzig}, in which the contribution to the free energy from the long-range  forces is expressed in powers of the inverse temperature, the reference high-temperature system being a system with infinitely strong, purely repulsive forces. The success of this perturbation expansion was demonstrated by Longuet-Higgens and Widom \cite{lw},  Barker and Henderson \cite{bh}, and soon after by Weeks, Chandler, and Andersen (WCA) in their seminal paper ``Role of repulsive forces in determining the equilibrium structure of simple liquids" \cite{wca}.

In perturbation theory a fundamental problem is how to separate the strong repulsive from the weaker, long-range attractions. One possible separation was proposed by Barker and Henderson, who marked the separation at the distance where the potential is zero. WCA demonstrated, however, that by choosing instead the separation at the potential minimum one obtains a much better agreement between the particle distributions of system and reference system. Doing so separates the forces into purely repulsive and purely attractive forces, and one may say that the original idea of van der Waals is here captured in its purest form. Numerous refinements of perturbation theory have since appeared; of particular relevance to the findings reported below are the works of Ree {\it et al.} from 1985 \cite{Ree} and of Hall and Wolynes from 2008 \cite{Hal08}, who studied the effects of a shifted-forces cutoff placed at a distance that scales with density in the same way as the radius of the FCS. An excellent summary of perturbation approaches up until 1976 can be found in Barker and Henderson's classic review \cite{BHrmp}.

The present work is motivated by two recent papers of Berthier and Tarjus. In 2009 they showed \cite{ber09} that the WCA approximation gives much too fast dynamics for the Kob-Andersen binary Lennard-Jones (KABLJ) \cite{KA} viscous liquid. They concluded that, while the attractive forces have little effect on the liquid's structure, these forces affect the dynamics in a ``highly nontrivial and nonperturbative way''. This led Pedersen {\it et al.} to investigate whether it is possible to reproduce both statics and dynamics of the KABLJ liquid by purely repulsive potentials different from the WCA potentials \cite{ped10}. This is indeed possible using inverse power-law (IPL) potentials; even the thermodynamics is slightly better predicted using IPL potentials than using the WCA approximation to the true LJ potentials \cite{ped10}. Altogether, this indicates that it is not the presence of  attractive forces {\it per se}, which is responsible for getting the correct dynamics, leaving open the question why the WCA approximation fails so dramatically for the KABLJ liquid's viscous dynamics. Very recently the problem was reconsidered by Berthier and Tarjus \cite{ber11}. From a careful numerical investigation of the standard LJ and the KABLJ liquids they conclude that in the viscous regime the WCA approximation gives results that are quantitatively and qualitatively different from the true potentials, a finding ``which appears to contradict the common view that the physics of dense liquids is dominated by the steep repulsive forces''. Moreover, they find that ``a key aspect in explaining the differences in the dynamical behavior of the two models [WCA vs. true potential] is the truncation of the interaction potential beyond a cutoff''.

This last point, in conjunction with recent results of ours on the shifted-forces cutoff \cite{toxdyre}, inspired to the present investigation that systematically studies the effect of neglecting the forces beyond a shifted-forces cutoff. We conclude below that the crucial point is to include all interactions from the FCS; the effect of interactions beyond the FCS on the thermodynamics can be taken into account to a good approximation by first-order perturbation theory. 

\begin{figure}
\begin{center}
\includegraphics[width=6cm,angle=-90]{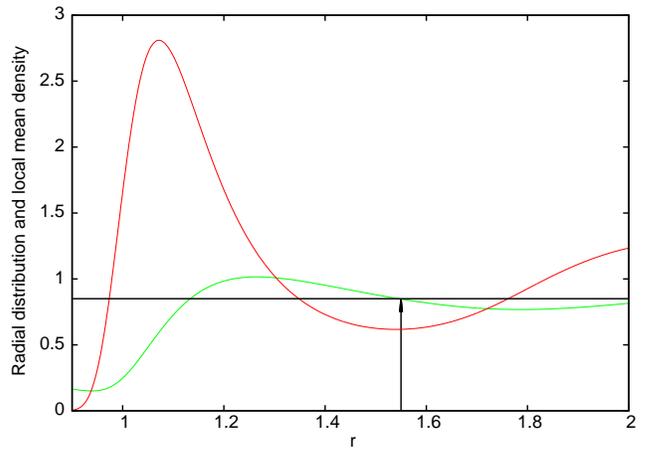}
\caption{ Radial distribution and local mean density in a condensed LJ fluid at temperature $T=1.00$ and density $\rho=0.85$ (black straight line). With red $g_{\textrm{LJ}}(r)$, with green the mean density within a sphere with radius $r$ and a particle at its center. The minimum of $g_{\textrm{LJ}}(r)$ at $r=1.55 \sigma$ indicated by an arrow (corresponding to 5.28\AA \ for Ar) coincides with the radius for which the local density equals the mean density $\rho=0.85$.
\label{vdW}}
\end{center}
\end{figure}

The radius of the FCS for a fluid of simple spherically symmetrical molecules can either be defined as the distance where the radial distribution function has its first minimum or, following van der Waals \cite{rowwidom}, as the distance from a particle for which the mean density within a sphere with this radius equals the overall particle density. For the LJ system these two distances are almost identical (Fig. \ref{vdW}); for the KABLJ system there is a small, but insignificant difference.

A condensed fluid may be thought of as similar to a distorted crystal with holes. Along this line of thinking, the importance of the FCS in condensed fluids was acknowledged long ago by  Frenkel \cite{Frenkel}, Eyring \cite{Eyring}, and Lennard-Jones and Devonshire \cite{Lennard-Jones,Devonshire}. Hole theories and lattice theories for liquids are discussed, e.g., in Refs. \onlinecite{Hirsfelder,Rowlinson}. The findings reported below confirm van der Waals' idea of the central role of the forces from particles within a local shell where the number density equals the overall density, as well as the basic ideas behind the hole and lattice theories for condensed fluids. 

Section II presents our results for the standard Lennard-Jones system, Sec. III investigates the highly viscous KABLJ mixture. Section IV summarizes briefly the consequences of our findings for the general picture of simple liquids and for the role of perturbation theory for understanding the physics of simple liquids.

\section{Simulations of the Lennard-Jones fluid with varying cutoffs}

The basic idea behind the standard perturbation expansion for fluids is that the particle distribution is determined primarily by the repulsive part of the potential. In this generally accepted picture the effect of the longer-range attractive part of the potential can be obtained from the first (mean-field) contribution in the expansion.

For a given state point with density $\rho$ and temperature $T$ the validity of the mean-field approximation may be tested by determining structure and dynamics of the fluid when the spatial range of the interactions is varied in the simulations via the cutoff. We make the model as simple as possible. The basic pair potential is the Lennard-Jones (LJ) function

\begin{equation}
u(r)=4 \epsilon [(r/\sigma)^{-12} -(r/\sigma)^{-6}]\,.
\end{equation}
The range of interactions is limited by introducing a shifted-forces (SF) cutoff at a radius denoted by $r_c$ \cite{Tildesley,toxdyre}. A SF cutoff  has the pair force go continuously to zero at $r_c$, which is obtained by subtracting a constant term as follows 

\begin{equation}\label{shf}
f_{\rm SF}(r)\,=\,
\begin{cases}
f_{\rm LJ}(r) - f_{\rm LJ}(r_c) & \text{if}\,\,  r<r_c \\
0 &  \text{if}\,\, r>r_c\,\,.
\end{cases}
\end{equation}
This corresponds to the following modification of the potential: $u_{\rm SF}(r)=u_{\rm LJ}(r) -(r-r_c) u'_{\rm LJ}(r_c)-u_{\rm LJ}(r_c)$ for $r<r_c$, $u_{\rm SF}(r)=0$ for $r>r_c$. Reference \onlinecite{toxdyre} showed that a SF cutoff gives more accurate results than the traditionally used shifted-potential (SP) cutoff \cite{comment}.  

Neglecting the forces beyond $r_c$ leads to a difference in the thermodynamics, dynamics, and structure of the system. After introducing a cutoff the pair distribution function $g(r)$ deviates from the ``true'' $g_{\textrm{LJ}}(r)$. In the mean-field approximation, the energy and pressure include the effect of the neglected forces as a mean contribution, calculated from the particle distribution of the reference system (which ignores the long-range forces). For the pressure, $p(\rho, T)$, the mean-field approximation is \cite{bh}

\begin{equation}
 p(\rho, T) =  \rho T   +\frac{2 \pi}{3} \rho^2 \int_{0}^{\infty}  g(r)f_{\textrm{LJ}}(r)r^3 dr.
\end{equation}
Any difference between the mean-field approximation of, e.g., the pressure and the true value is caused by a different particle distribution originating from the exclusion of the forces beyond $r_c$. By varying the cutoff we can thus test the validity of the mean-field approximation. The validity of the first-order perturbation expansion may be investigated by determining the difference between the (radial) distributions

\begin{equation}\label{4}
 \Delta g(r) =g(r)-g_{\textrm{LJ}}(r).
\end{equation}
The potential energy and the pressure deviate from the correct values (denoted by $U_{\textrm{LJ}}$ and $p_{\textrm{LJ}}$) by the terms

\begin{equation}
\Delta  U(\rho, T)/N = 2 \pi \rho \int_{0}^{\infty} \Delta g(r)u_{\textrm{LJ}}(r)r^2 dr, 
\end{equation}
and

\begin{equation}\label{6}
\Delta p(\rho, T) = \frac{2 \pi}{3} \rho^2 \int_{0}^{\infty} \Delta g(r)f_{\textrm{LJ}}(r)r^3 dr,
\end{equation}
In molecular dynamics these integrals are obtained directly as time and particle averages of the explicit interaction energies, $u(r_{i,j})$, and forces, $f(r_{i,j})$, between pairs of particles, $i$ and $j$. In standard perturbation theory to first order the Helmholtz free energy difference between the system with a cutoff and the true LJ system is given by 

\begin{equation}
\Delta  F(\rho, T) = \Delta  U(\rho, T)\,.
\end{equation}
The perturbation expansion is asymptotically correct at high temperatures for systems with  infinitely strong purely repulsive forces \cite{Zwanzig}.i

Five  state points of the standard single-component LJ condensed fluid were investigated \cite{toxmd}. The five state points are: 

\begin{enumerate}
\item $\rho$ = 0.85, $T$ = 1.00 (liquid at medium temperature);
\item $\rho$ = 0.85,  $T$  = 0.65 (liquid at low temperature and almost zero pressure);
\item $\rho$ = 2.5,  $T$  = 100 (condensed fluid at very high  density and temperature);
\item $\rho$ = 0.60,  $T$  = 1.5 (low-density condensed fluid just above the critical temperature);
\item $\rho$ = 1.05, T = 1.00 (fcc solid).
\end{enumerate}
The fluid systems studied consisted of $N = 2000$ particles, the fcc solid system of $N =12^3= 1728$ particles (using always periodic boundary conditions). For each value of the cutoff the systems were simulated for at least 16 million time steps ($\approx 160 ns$ in Argon units). Below we present data for the fluid state points (1)-(4), for which results from simulations using various values of the cutoff are compared to results for the true LJ system, which is here defined by the cutoff $r_c=4.5 \sigma$ (these data are denoted by LJ). Results for the crystalline state point (5) confirm the general physical picture arrived at.

\subsection{Thermodynamics and dynamics}

\begin{figure}
\begin{center}
\includegraphics[width=6cm,angle=-90]{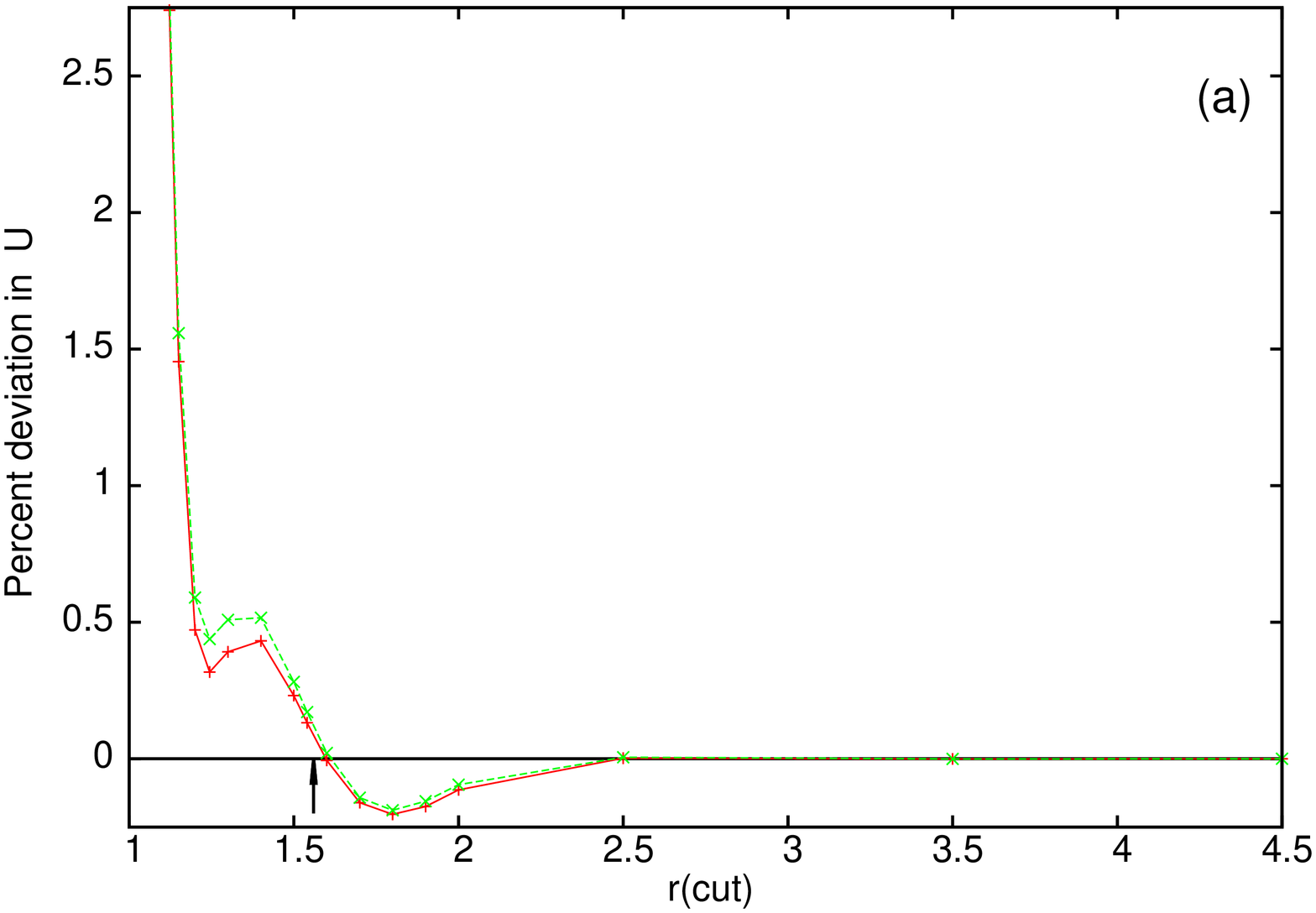}
\includegraphics[width=6cm,angle=-90]{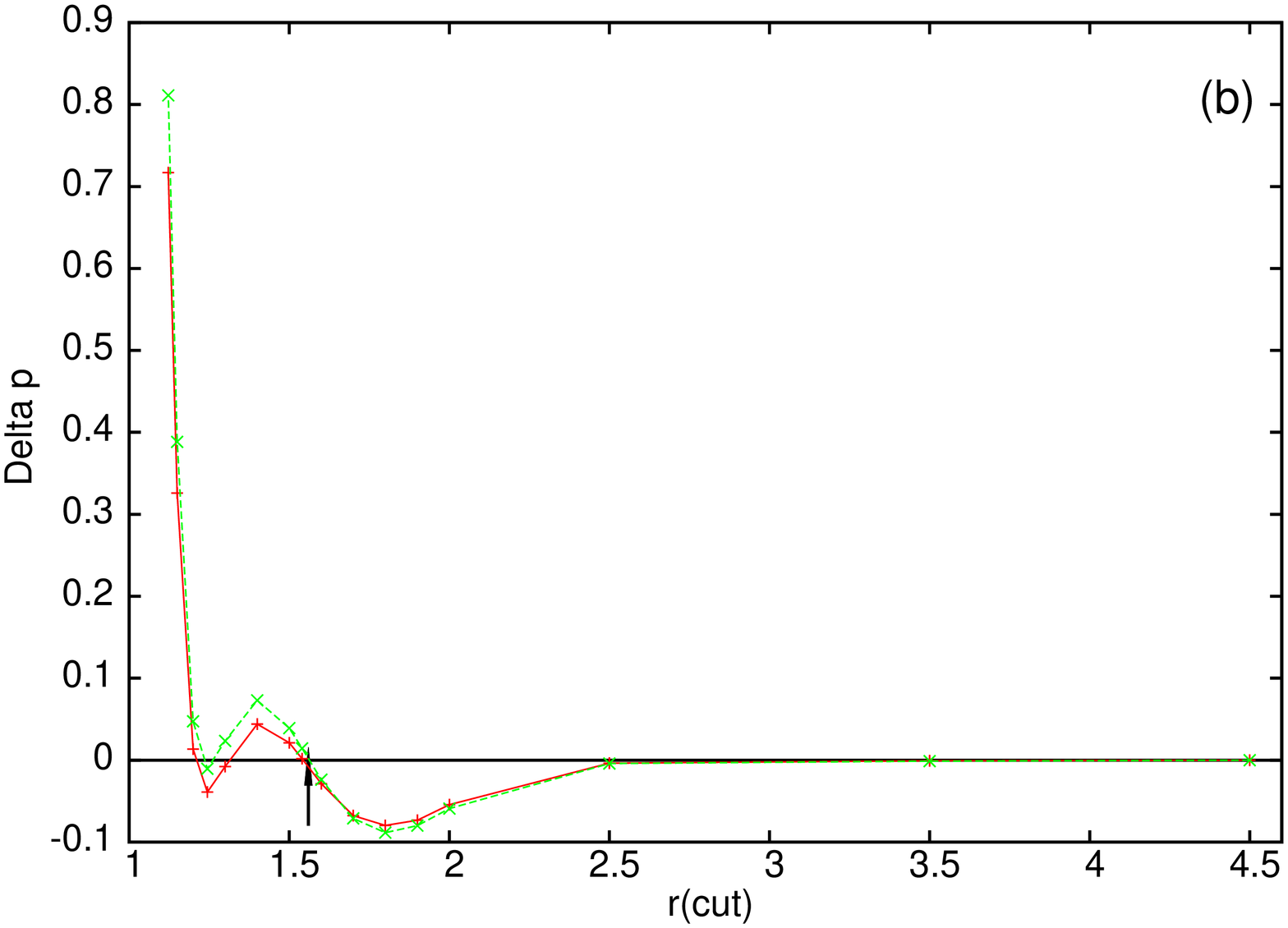}
\end{center}
\caption{Changes of potential energy $\Delta U$and pressure $\Delta p$ as functions of the cutoff (a shifted-forces cutoff is used throughout the paper).
In each figure the arrow marks the location of the first minimum of $g(r)$, which delimits the first coordination shell (FCS) (each figure should have two arrows, but they virtually coincide and only one is shown). 
(a) $\Delta U/U_{\textrm{LJ}}$ in percent as a function of the cutoff in the LJ fluid with density $\rho=0.85$. The data in red are for $T=1$ at which  $U/N=-5.7396$ (state point (1)), the data in green are for $T=0.65$ at which $U/N=-6.1242$  (state point (2)) \cite{explain}. 
(b) $\Delta p$ as a function of the cutoff in the  LJ fluid with density $\rho=0.85$. The data in red are for $T=1$  (state point (1)) where $p_{\textrm{LJ}}=2.0769$, the data in green are for $T=0.65$ where $p_{\textrm{LJ}}=-0.0912$  (state point (2)) \cite{explain}. 
\label{thermo}}
\end{figure}

\begin{figure}\begin{center} 
\includegraphics[width=6cm,angle=-90]{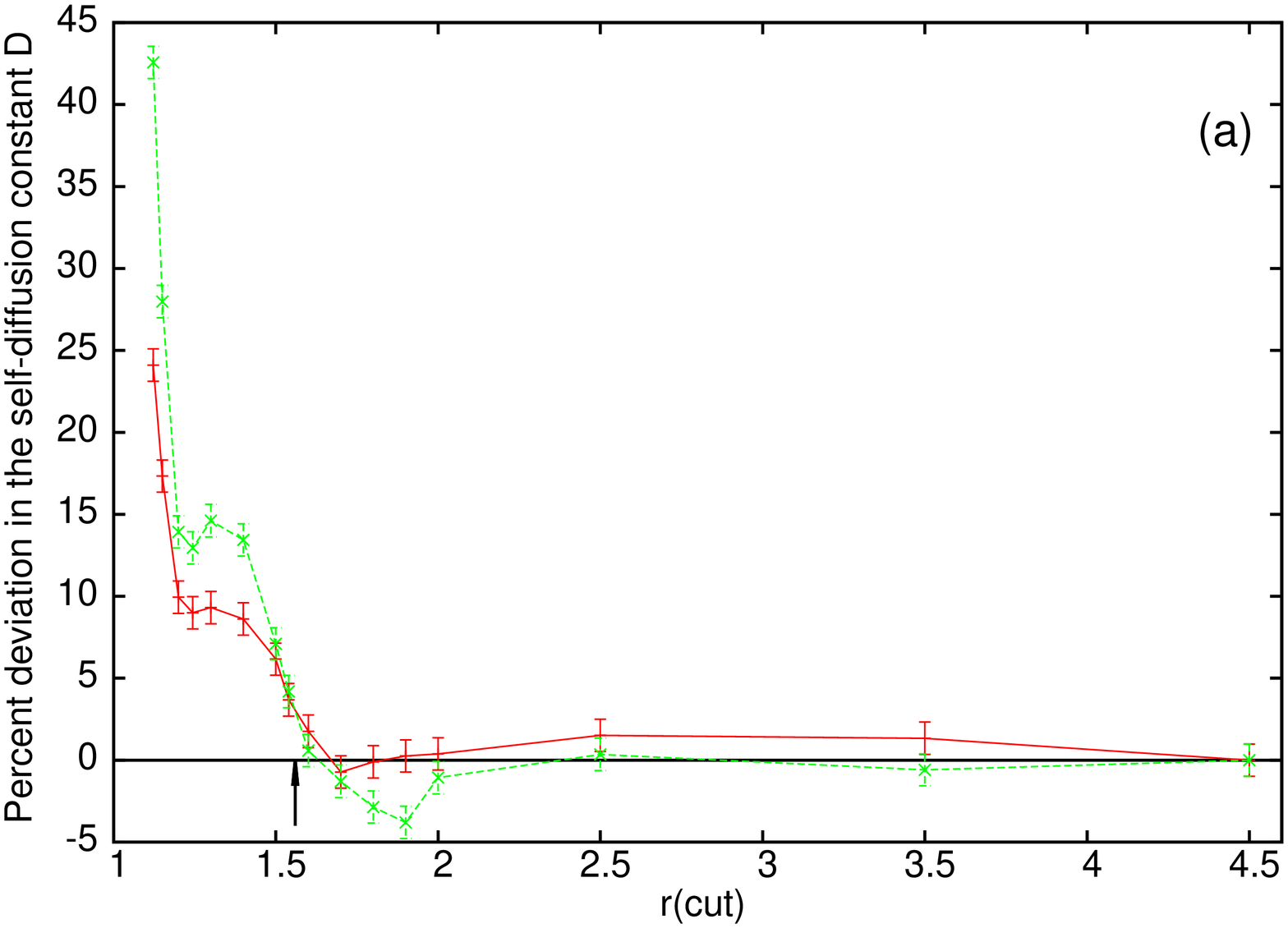}
\includegraphics[width=6cm,angle=-90]{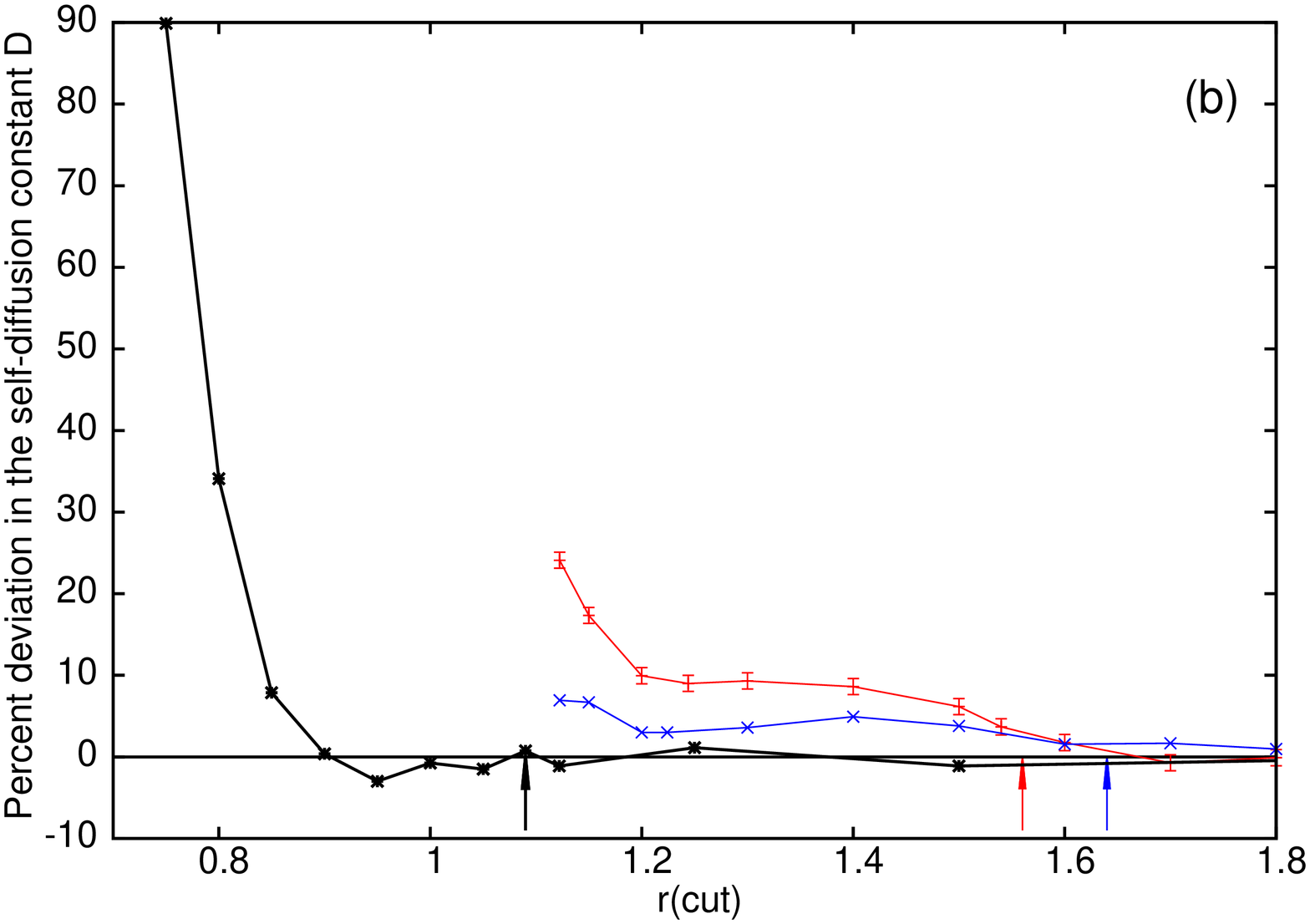}
\end{center}
\caption{Relative changes of the self-diffusion constant $\Delta D$ at various cutoffs.
(a) $\Delta D/\Delta D_{\textrm{LJ}}$ in percent as a function of the cutoff in a LJ fluid system at the density $\rho=0.85$. The data in red are for $T=1$ where $D_{\textrm{LJ}}= 0.0497$, the data in green are for $T=0.65$ where $D_{\textrm{LJ}}= 0.0258$ (the arrow marks the location of the first minimum of $g(r)$).
(b) $\Delta D/\Delta D_{\textrm{LJ}}$ for different densities. The red curve is for state point (1), the black curve is for the high-density, high-temperature state point (3) ($\rho,T=2.5,100$). The blue curve is for state point (4) ($\rho,T=0.60,1.5$). The three arrows mark the locations of the first minima of $g(r)$ for the respective state points.\label{dynamics}}
\end{figure}

Figure \ref{thermo}(a) shows the change in percent of the potential energy as a function of the cutoff obtained by Eq. (5) at state points (1) (red) and (2) (green), Fig. \ref{thermo}(b) shows the change in pressure using Eq. (6)  for the same simulations. Figure \ref{dynamics}(a) shows the difference in self-diffusion constant $\Delta D$ for the LJ fluid at state points (1) (red) and (2) (green), Fig. \ref{dynamics}(b) shows the same for state points (1), (3), and (4). The arrows mark the locations of the first minimum of the radial distribution function, delimiting the FCS. 

For both thermodynamics and dynamics we find the following: There is virtually no effect on structure and dynamics of ignoring the forces beyond the standard cutoff used in most LJ simulations, $r_c=2.5 \sigma$. This is not surprising in view of the fact that the interactions are very small at such large distances. There is a small effect on the dynamics of neglecting the forces from distances $1.5 \sigma \le 2.5 \sigma$. Interestingly, the accuracy of the simulations  ``recovers'' and the simulations generally agree well with those of the true LJ system when $r_c \approx 1.5 \sigma$. This is where $g_{\textrm{LJ}}(r)$ has its first minimum (marked by arrows in Figs. \ref{thermo} and \ref{dynamics}), delimiting the first coordination shell (FCS). As forces are gradually removed for $r< 1.5 \sigma$, however, both thermodynamics and dynamics begin to deviate significantly from those of the true LJ system. In particular, the self-diffusion constant increases significantly. Note that using the WCA cutoff $r_c=2^{1/6}$ (at the potential minimum), the energy obtained by the mean-field approximation deviates only a few percents from the correct value. This indicates that even modifications of the potential that lead to only quite small changes in free energy (Eq. (7)) may significantly affect the dynamics. 

To summarize the findings so far: 1) Simulations of the LJ liquid's thermodynamics and dynamics are fairly reliable when the cutoff is placed at $g(r)$'s first minimum; 2) the small energy changes -- and thereby small free energy changes -- induced by removing even a minor part of the attractive forces within the FCS lead to a significant increase of the self-diffusion constant. This suggests the following. For a condensed, uniform fluid the forces on a given particle from particles within its entire FCS play crucial roles for the dynamics, whereas the forces from particles outside the FCS have little influence and their contribution to the thermodynamics may be taken into account to a good accuracy via mean-field methods. There is a small effect on the thermodynamics and, in particular, on  the self-diffusion constant by also including  the forces just outside  the FCS: by including the attractive forces from the interval $1.5 \sigma \le r\le  2.0 \sigma$ the energy decreases by one percent, whereas these attractive forces lower the self-diffusion constant by up to four percent for state point (2).

With the van der Waals picture and perturbation theory in mind the following question now arises: Is it important that the forces within the first coordination shell includes some attractions? The answer to this question is no. Recall that a purely repulsive inverse power-law system does exist that has almost the same radial distribution function as the LJ system \cite{Bailey,Dyre4}. Since the potential of the mean force $w(r)$ on a particle is given by the radial distribution function \cite{chandler}, $w(r) = -k_BT\ln[g(r)]$, the mean force on a LJ particle from the particles in its FCS does not depend directly on the sign of these forces but merely on $g(r)$.

In order to demonstrate that the dynamics of a LJ fluid is obtained correctly even when only purely repulsive parts of the LJ potential lie within the FCS, we simulated the LJ system at such a high density that the entire FCS is within the ``WCA range'' delimiting the range of the repulsive forces ($r<2^{1/6}$): $\rho=2.5$, state point (2). Even at this state point a cutoff at $g_{\textrm{LJ}}(r)$'s first minimum ($r_c=1.09 \sigma$) gives good results for self diffusion (Fig. \ref{dynamics}(b)). This shows that inquiring into the respective roles of repulsive versus attractive forces is less central than focusing on the role of the FCS for simple condensed liquids' thermodynamics and dynamics -- the main conclusion of the present paper. Figure \ref{dynamics}(b) shows the relative deviation of the self-diffusion constant $D$  for state points (1) and (3), demonstrating that this behavior is independent of whether forces are attractive or repulsive. Included in this figure is also the relative change in self-diffusion constant for a LJ fluid at state point (4) (blue dashes). This low-density state point does not really have a well-established coordination shell, and the diffusion does not show the same strong dependence of the forces within the first shell of nearest neighbour; nevertheless, this state point shows the same trend as the other state points.

\begin{figure}\begin{center}
\includegraphics[width=6cm,angle=-90]{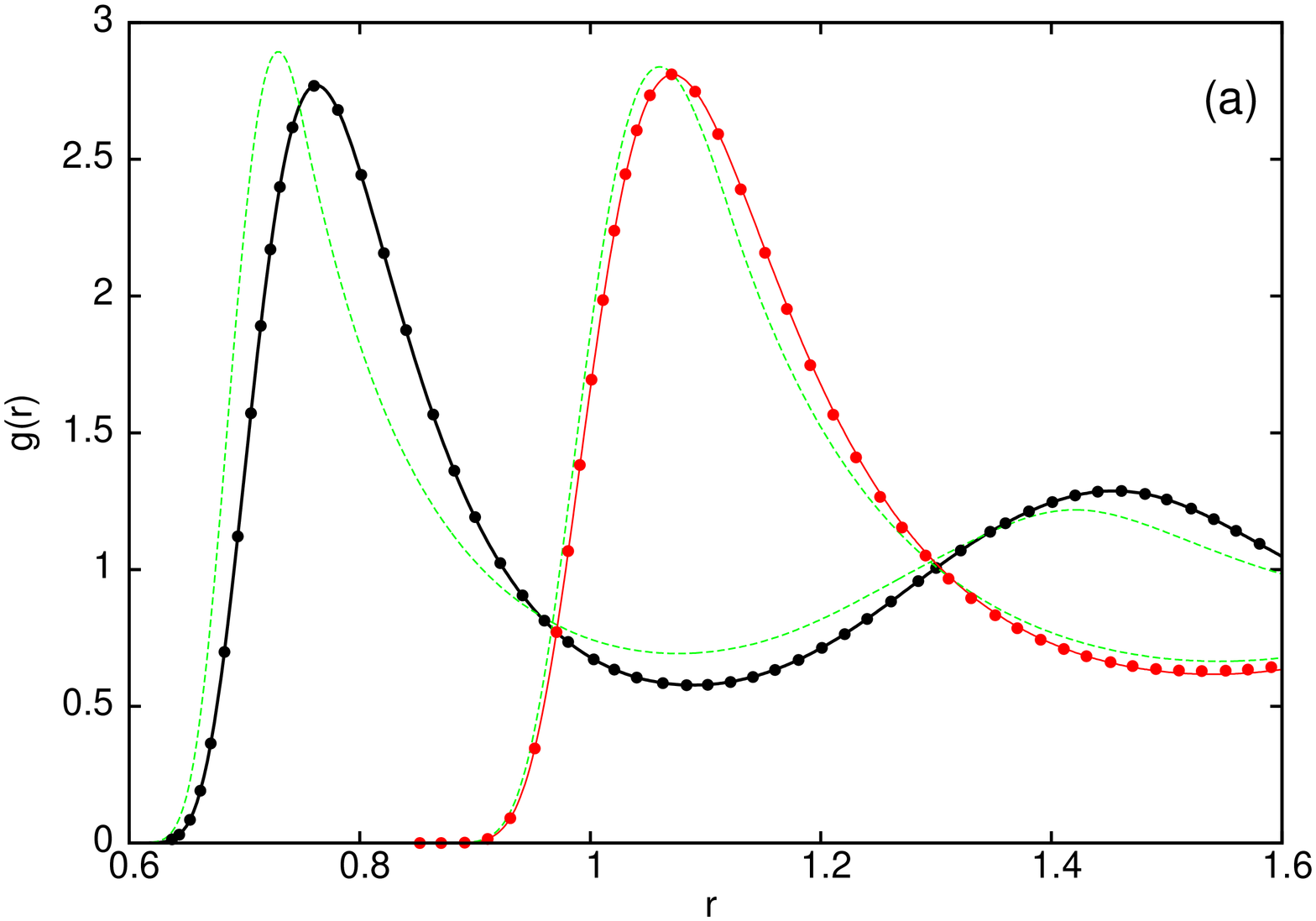}
\includegraphics[width=6cm,angle=-90]{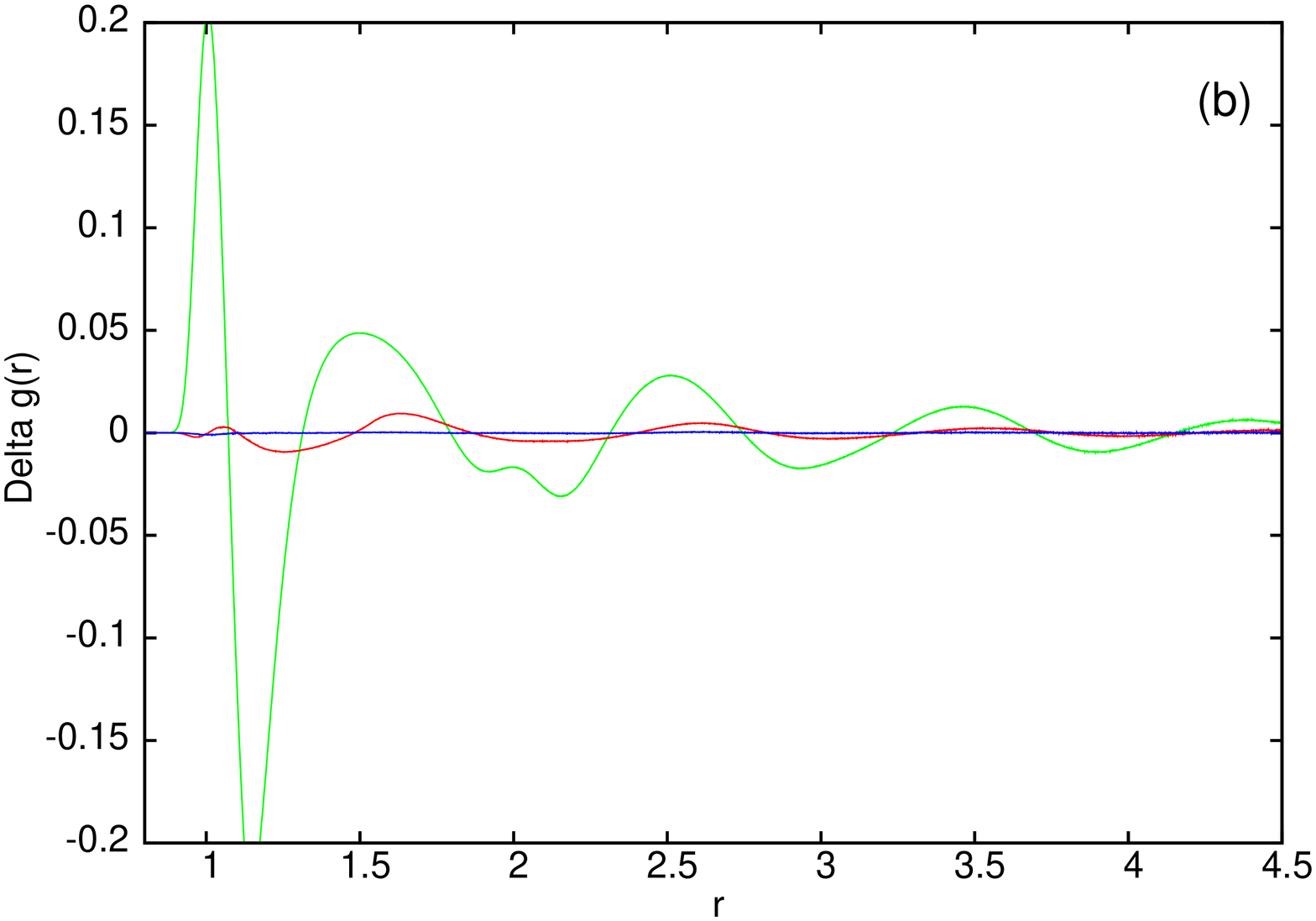}
\end{center}
\caption{(a) Radial distribution functions for the two state points (1) ($(\rho,T)=(0.85,1.00)$) and (3) ($(\rho,T)=(2.5,100)$) for different cutoffs. The black line gives $g_{\textrm{LJ}}(r)$ at state point (3), the red line gives $g_{\textrm{LJ}}(r)$ at state point (1). The solid  points give $g(r)$  at (1) with $r_c=1.5 \sigma$ and at (3) with  $r_c=1.09 \sigma$, respectively, i.e., for a simulation that includes only forces from particles within the FCS. The dashed green curves give $g(r)$ for (1) with $r_c=2^{1/6} \sigma$ and (3) with $r_c=0.75 \sigma$, respectively, i.e., keeping only forces up until the maximum of $g(r)$. (b) $\Delta g(r)$ for different values of the cutoff at state point (1). With green is shown results for a (WCA) cutoff at the minimum of the LJ potential ($r_c=2^{1/6} \sigma$). The red line is for a cutoff at the first minimum of $g_{\textrm{LJ}}= 1.55 \sigma$, the blue line is for $r_c=2.5 \sigma $. \label{structure}}
\end{figure}

\subsection{Structure}

Figure \ref{structure}(a) shows $g_{\textrm{LJ}}(r)$ at the high-density, high-temperature state point (3) (black) along with the corresponding distribution at state point (1) (red). The $g_{\textrm{LJ}}(r)$ of the high-density state point (3) is almost a simple scaling of $g_{\textrm{LJ}}(r)$ of state point (1) to a shorter distance \cite{Dyre4}, with maximum shifted from $r \approx 2^{1/6} \sigma$ at state point (1) to  $r \approx 0.75 \sigma$ at state point (3). As mentioned already, the first minimum in $g_{\textrm{LJ}}(r)$ at state point (1) is shifted from $r \approx 1.5 \sigma$ to a value at state point (3) of $r \approx 1.09 \sigma$, i.e., to a distance where all the particles within the FCS interact with the central particles by purely repulsive forces. Shown with solid  points are values of $g(r)$  for the systems where the forces beyond the FCS at state points (1) and (3) have been neglected. The two sets of distributions (line: $g_{\textrm{LJ}}(r)$ and dots: $g(r)$) are practically identical. This demonstrates that forces beyond the FCS play a negligible role for the distribution, no matter whether these forces are attractive or repulsive. The green dashes show the corresponding  distributions obtained when only the forces for distances shorter than the distance at the first maximum of  $g_{\textrm{LJ}}(r)$ are taken into account in the dynamics, i.e., using for state point (1) $r_c= 2^{1/6} \sigma$ and for state point (3) $r_c=0.75 \sigma$.

In perturbation theory the (radial) distribution of particles around a target particle is to a good approximation given by the distribution obtained from the short-range part of the pair interactions \cite{BHrmp}. For "WCA" perturbation theory the unperturbed system neglects all attractive interactions. Although the two distributions (green dashes and red lines of Fig. \ref{structure}(a)) agree reasonably well and lead to a fairly good thermodynamic description using the first-order perturbation corrections to the thermodynamic functions, there are systematic differences. To some extent, by neglecting the attractive forces in the dynamics there is a cancellation of errors in the radial distribution function in the integrals of  Eqs. (\ref{4})-(\ref{6}). Figure \ref{structure}(b) highlights $\Delta g(r)$ at state point (1) for three values of $r_{c}$. With green line is shown the differences for the WCA cutoff (purely repulsive forces), the red line gives $\Delta g(r)$ after including all forces within the FCS (the corresponding $g(r)$ is shown in the right part of Fig. \ref{structure}(a) with green dashes and red point, respectively). For reference, the blue line gives $\Delta g(r)$ for the standard shifted-potential cutoff at 2.5$\sigma$. The figure demonstrates that substantial improvement is obtained by including all particles within the FCS in the dynamics, although there are still small differences. The effect of these small differences cancels, however, to a much larger extent than when using the purely repulsive WCA reference system, and it leads to an excellent overall agreement  for both thermodynamics and dynamics (Figs. \ref{thermo} and \ref{dynamics}).

\begin{figure}\begin{center}
\includegraphics[width=6cm,angle=-90]{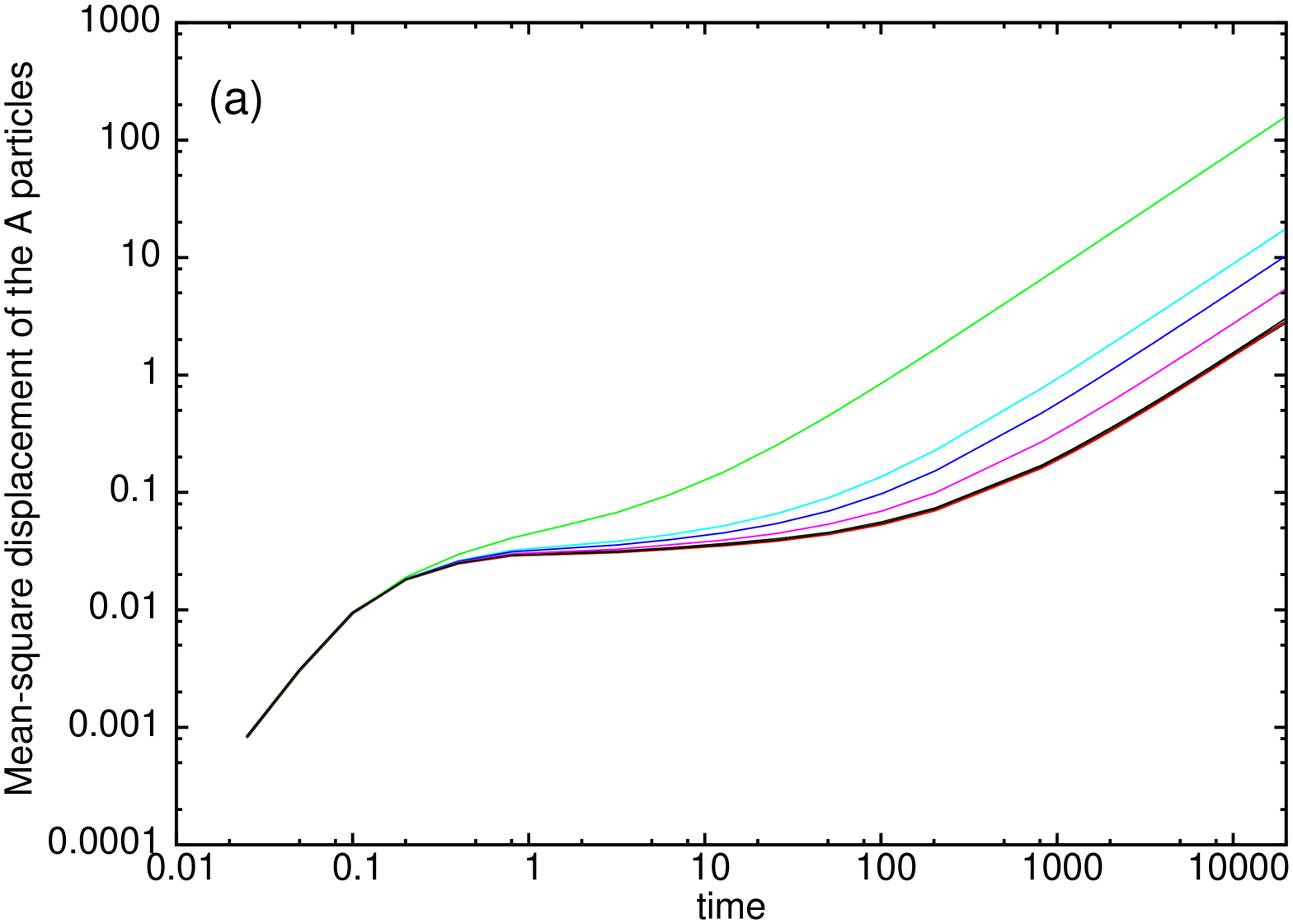}
\includegraphics[width=6cm,angle=-90]{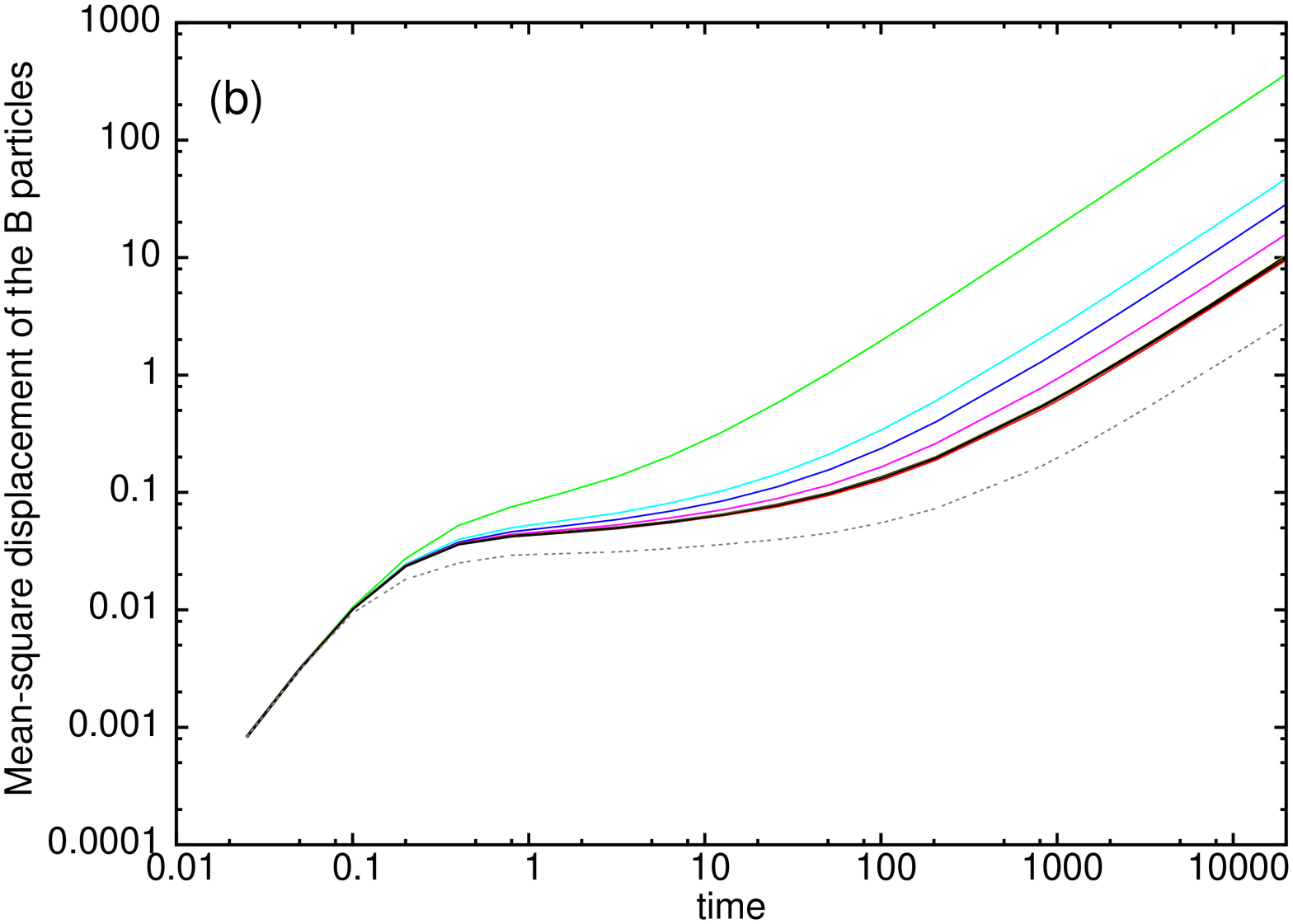}
\end{center}
\caption{
(a) Log-log plot of the mean-square displacement (msd) of the A particles in the KABLJ fluid as a function of time for different values of $r_c$. With black is shown the ``true'' msd for $r_c= 4.5 \sigma_{\textrm{AA}}$; green: $r_c= 1.12 \sigma_{\alpha, \beta}$ \cite{kacut};  light blue: $r_c= 1.2 \sigma_{\textrm{AA}}$; blue:  $r_c= 1.3 \sigma_{\textrm{AA}}$; magenta:  $r_c= 1.4 \sigma_{\textrm{AA}}$; red: $r_c= 1.5 \sigma_{\textrm{AA}}$; yellow: $r_c= 2.0 \sigma_{\textrm{AA}}$ and dark blue: $r_c= 2.5 \sigma_{\textrm{AA}}$ (there are no visible differences for $r_c \geq  1.5 \sigma_{\textrm{AA}}$).
(b) Log-log plot of the msd of the B particles in the KABLJ fluid as a function of time for different values of $r_c$. With black is shown the ``true'' msd for $r_c= 4.5 \sigma_{\textrm{AA}}$; green: $r_c= 1.12 \sigma_{\alpha, \beta}$ \cite{kacut};  light blue: $r_c= 1.2 \sigma_{\textrm{AA}}$; blue:  $r_c= 1.3 \sigma_{\textrm{AA}}$; magenta:  $r_c= 1.4 \sigma_{\textrm{AA}}$; red: $r_c= 1.5 \sigma_{\textrm{AA}}$; yellow: $r_c= 2.0 \sigma_{\textrm{AA}}$ and dark blue: $r_c= 2.5 \sigma_{\textrm{AA}}$ (there are no visible differences for $r_c \geq  1.5 \sigma_{\textrm{AA}}$). Dark small dashes: the msd of the A particles for $r_c= 4.5 \sigma_{\textrm{AA}}$, also shown in (a) with black.\label{msd}}
\end{figure}

\section {Results for the Kob-Andersen binary Lennard-Jones viscous liquid}

The previous section showed that what matters for simulating with high accuracy the condensed LJ liquid's thermodynamics, dynamics, and structure, is to ensure that all interactions within the FCS are included in the dynamics -- independent of  whether or not there are any attractive forces coming from particles within the FCS. We moreover showed that accurate results are obtained by placing the cutoff right at $g(r)$'s first minimum, which provides a convenient and obvious definition of the border of the FCS. The 2009 paper by Berthier and Tarjus \cite{ber09}, which reopened the discussion of the role of the attractive forces in simulations, focused on the highly viscous Kob-Andersen binary Lennard-Jones (KABLJ) liquid \cite{KA}. For this system the WCA cutoff leads to a dynamics that is up to two orders of magnitude too fast. Thus the highly viscous KABLJ system is particularly sensitive, and it is therefore important to test whether the dominance of the forces coming from particles within the FCS identified in the LJ simulations applies also for the KABLJ liquid.

The KABLJ liquid is a mixture of 80\% large (A) and 20\% small (B) LJ particles ($\sigma_{\rm BB}=0.88\sigma_{\rm AA}$) at the density $\rho \sigma_{\rm AA}^3$=1.2 and with a very strong AB attraction \cite{KA}. The A particles dominate the overall dynamics, and the (few and small) B particles are to a large extent slaves of the structure set by the A particles. For this reason, when we refer below to the FCS of the KABLJ system, it means always the first coordination shell of the A particles delimited by the first minimum of $g_{\rm AA}(r)$. Likewise, cutoffs of all interactions are given in units referring to $\sigma_{\rm AA}$ \cite{kacut}.

\begin{figure}\begin{center}
\includegraphics[width=6cm,angle=-90]{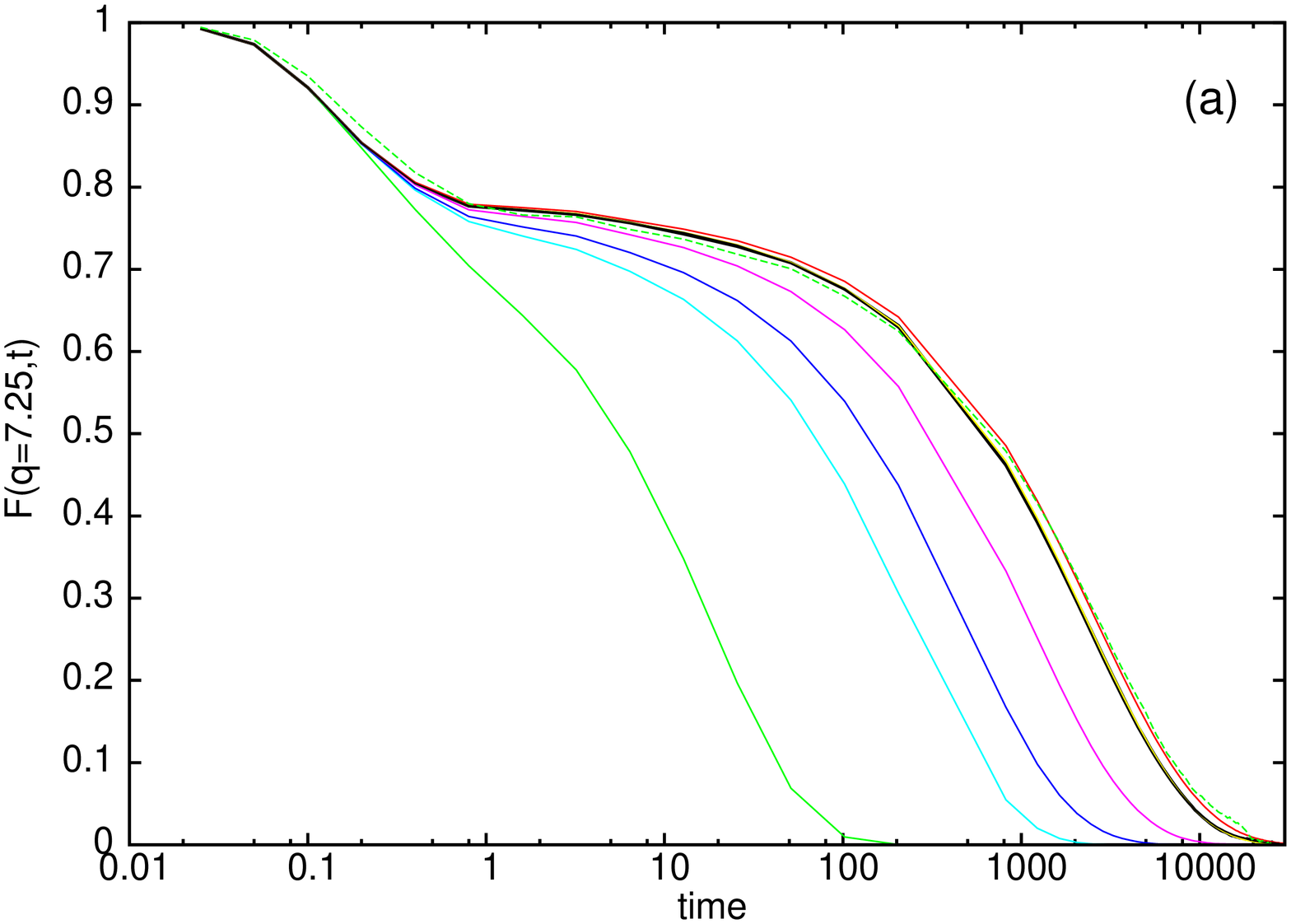}
\includegraphics[width=6cm,angle=-90]{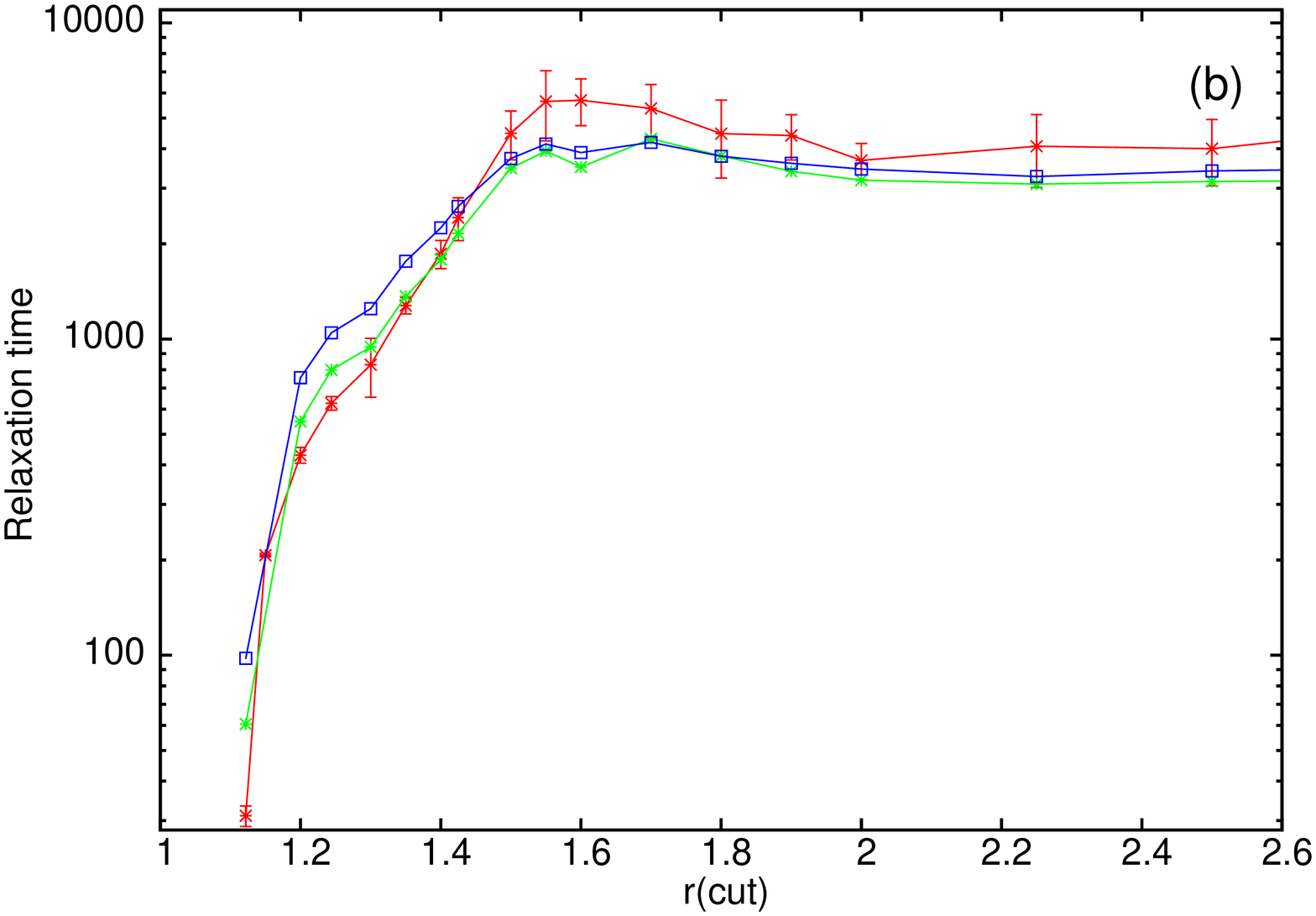}
\end{center}
\caption{ 
(a) Incoherent intermediate scattering function $F_s(q,t)$ at the wave vector $q \sigma_{\textrm{AA}}=7.25$ for the A particles in the KABLJ fluid at $T=0.45$ and density $\rho \sigma_{\rm AA}^3$=1.2 as a function of time for different values of $r_c$. With black is shown $F(q,t)$ for $r_c= 4.5 \sigma_{\textrm{AA}}$. Green: $r_c= 1.12 \sigma_{\alpha, \beta}$ \cite{kacut}, the line is for
$T=0.45$ and green dashes is for $T=0.3125$;  light blue: $r_c= 1.2 \sigma_{\textrm{AA}}$; blue:  $r_c= 1.3 \sigma_{\textrm{AA}}$; magenta:  $r_c= 1.4 \sigma_{\textrm{AA}}$; red: $r_c= 1.5 \sigma_{\textrm{AA}}$; yellow: $r_c= 2.0 \sigma_{\textrm{AA}}$ and dark blue: $r_c= 2.5 \sigma_{\textrm{AA}}$ (the differences for $r_c \geq  1.5 \sigma_{\textrm{AA}}$ are within the uncertainties).
(b) The relaxation time $\tau$ of the KABLJ fluid at $T=0.45$. With read line and error bars are shown $\tau$ for the A particles obtained from Eq. (8). The self-diffusion constants $D_{\textrm{A}}^{-1}$ (green line) and $D_{\textrm{B}}^{-1}$ (blue line) are scaled to agree with $\tau$ for $r_c=4.5 \sigma_{\textrm{AA}}$.\label{KABLJdyn}} 
\end{figure}

The mean-square displacement of a particle in a viscous fluid separates into two regimes: the short-time ballistic regime in which the particle vibrates within its FCS, and the long-time diffusive regime that reflects the occasional escape from the shell.  The importance of the range of the forces for viscous fluids is investigated below by simulating the KABLJ system of  $N = 1000$ particles at $\rho \sigma_{\textrm{AA}}^3 =1.2 $ and $T$  = 0.45. In order to determine the time correlations in this highly viscous fluid state accurately the simulations were very long, typically 160 million time steps. 

The mean-square displacements (msd) for different cutoffs are shown as functions of time in Figs. \ref{msd}(a) and (b), which gives msds of the large (A) and the small (B) particles, respectively. Both figures show the same qualitative behavior. The displacements change drastically and systematically by increasing the cutoff  from $r_c=1.12  \sigma_{\alpha, \beta}$ to $r_c = 1.5 \sigma_{\textrm{AA}}$, but for $r_c \geq 1.5 \sigma_{\textrm{AA}}$ there are no visible differences -- here both A and B particles are trapped for a long time, which is independent of the long-range attraction.

The A particle self-intermediate scattering function at $q \sigma_{\textrm{AA}}=7.25$,

\begin{equation}
 F_s(q,t)=\frac{1}{N} \left \langle \sum_{j=1}^{N} e^{i \mathbf{q} \cdot (\mathbf{r}_j(t)-\mathbf{r}_j(0))} \right \rangle\,,
\end{equation}
is traditionally used for probing the dynamics of the KABLJ viscous liquid. This function is shown in Fig. \ref{KABLJdyn}(a) for various cutoffs. The conclusion is the same as for the behavior of the mean-square displacements: forces from particles beyond $r \approx 1.5 \sigma_{\textrm{AA}} $ play little role for $F_s(7.25,t)$. The two green curves in Fig. \ref{KABLJdyn}(a) are for a WCA cutoff ($r_c=2^{1/6} \sigma_{\alpha, \beta}$) \cite{kacut}. The green line is for $T=0.45$ and the green dashes (almost invisible) is for a lower temperature, $T=0.3125$. At this lower temperature the self-intermediate scattering function and the self-diffusion constant for the KABLJ-WCA system agree with those of the KABLJ system at $T=0.45$. The dynamics of the KABLJ-WCA mixture at $T=0.3125$  is indistinguishable from that of the KABLJ system at $T=0.45$, which shows that the two systems agree after a simple temperature shift. A similar scaling behavior has been reported for simple LJ fluids \cite{ya}. (However, it should be noted that the KABLJ-WCA system is more prone to crystallization because the exothermic binding between solute and solvent molecules is not present. It is possible to maintain the stability of this viscous mixture by keeping the exothermic attraction between solute and solvent particles \cite{toxcrys}.)   

Relaxation times were calculated from $F_s(7.25,t)$ via 

\begin{equation} 
 \tau= \int_0^\infty t  F_s(7.25,t) dt/\int_0^\infty   F_s(7.25,t) dt\,.
\end{equation}
Results are shown in  Fig. \ref{KABLJdyn}(b) along with those obtained from the self-diffusion via the slopes ($6D$) of the mean-square displacements using $\tau(D)\propto a^2/D$, where the relaxation times $\tau(D)_{\textrm{A}}$ and $\tau(D)_{\textrm{B}}$ were scaled to agree for $r_c=4.5 \sigma_{\textrm{AA}}$. Within the statistical uncertainties the three relaxation times agree and show the same dependence on $r_c$. These findings for the dynamics of the highly viscous KABLJ fluid confirm that it is the forces from distances below $1.5 \sigma_{\textrm{AA}}$ that give the correct high viscosity and long relaxation time of the viscous state. 

\begin{figure}\begin{center}
\includegraphics[width=6cm,angle=-90]{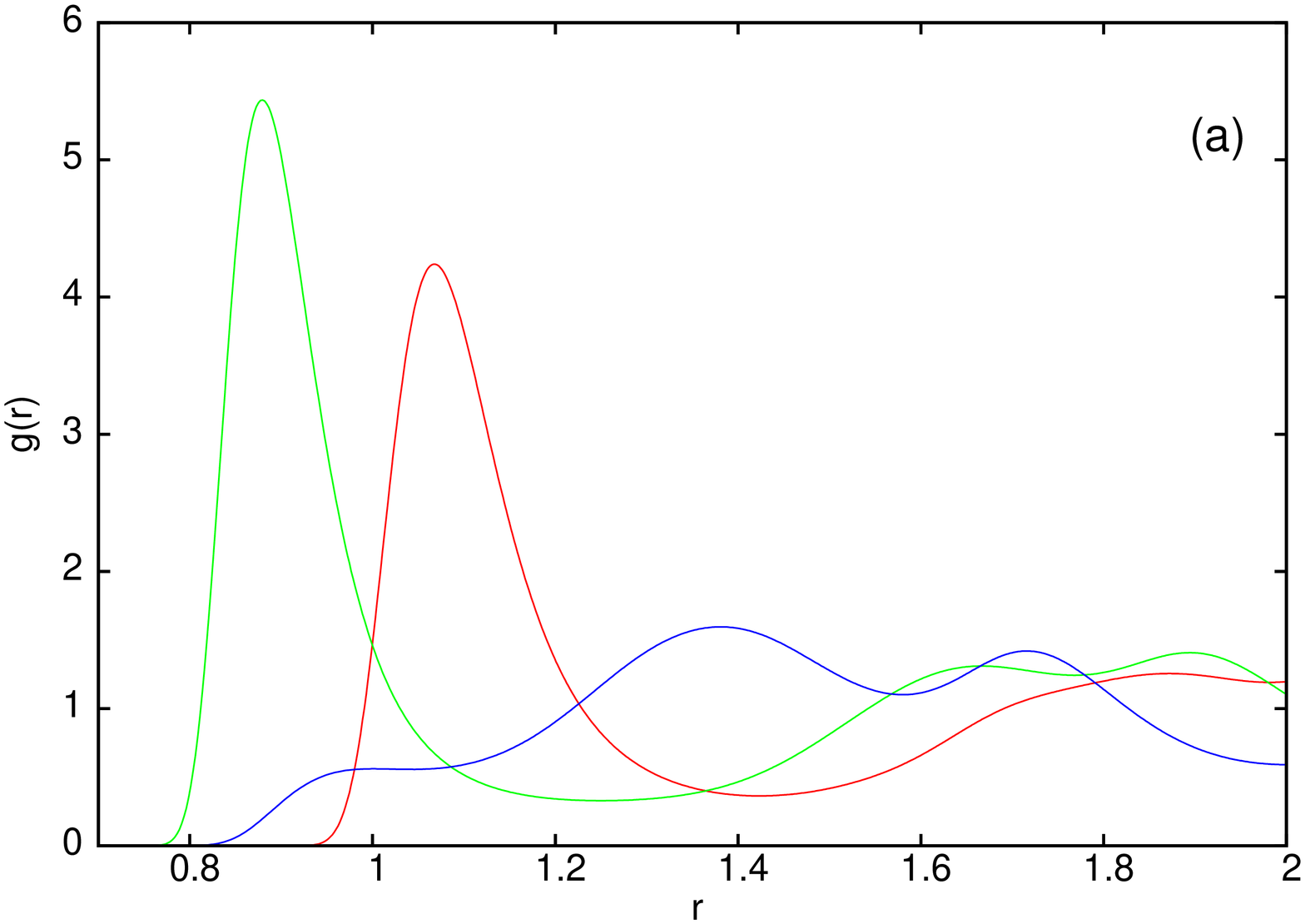}
\includegraphics[width=6cm,angle=-90]{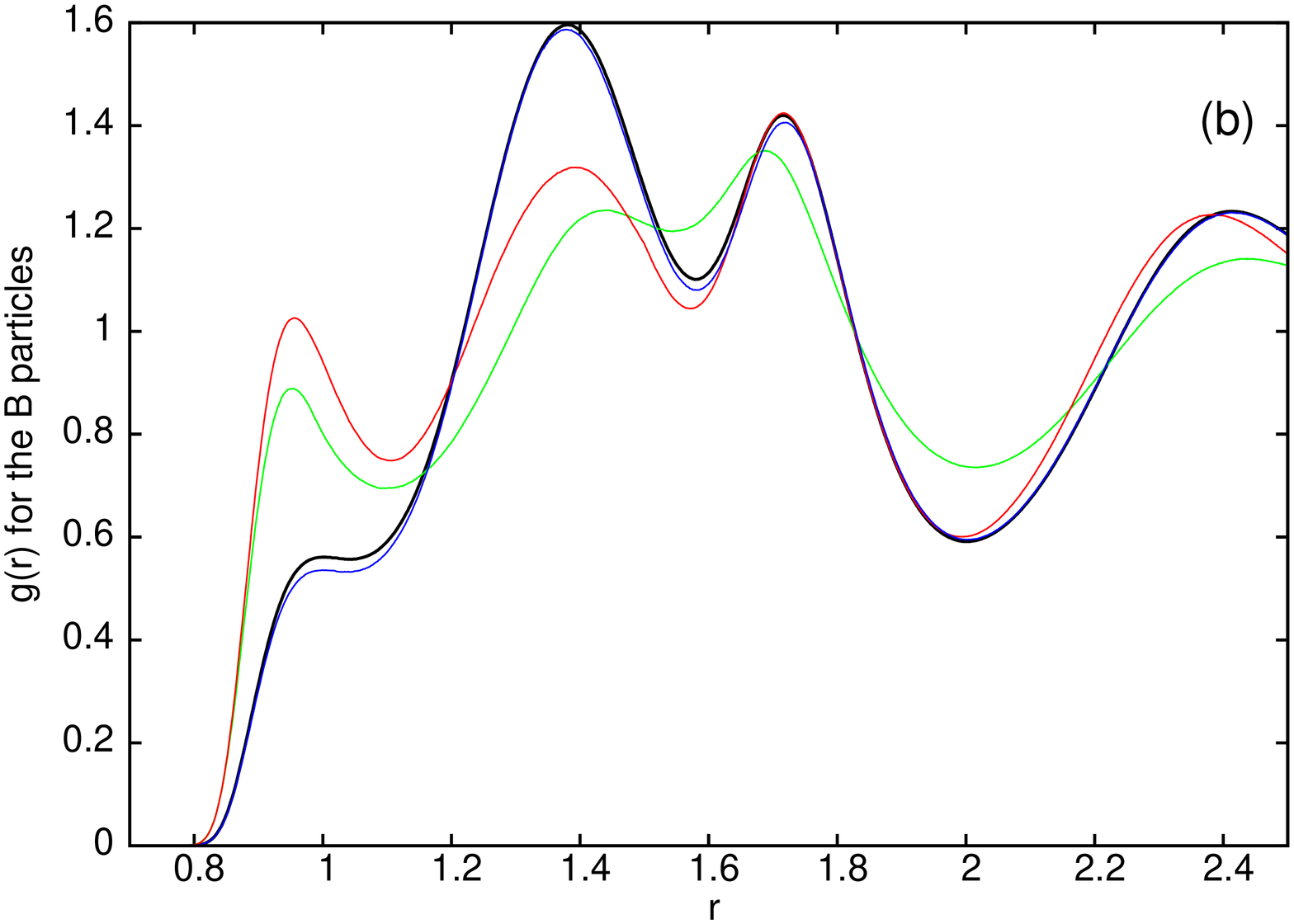}
\end{center}
\caption{(a) Radial distribution functions, $g_{\alpha,\beta}(r)$ for the KABLJ fluid at $T=0.45$. Red line: $g_{\textrm{AA}}(r)$;  green: $g_{\textrm{AB}}(r)$ and blue: $g_{\textrm{BB}}(r)$.
(b ) Radial distribution functions, $g_{\textrm{BB}}(r)$ for the KABLJ fluid $T=0.45$. Black line: $g_{\textrm{BB}}(r)$ for  $r_c =4.5 \sigma_{\textrm{AA}}$; green line: $g_{\textrm{BB}}(r)$ for $r_c =1.12 \sigma_{\alpha, \beta}$ \cite{kacut}; red line: $g_{\textrm{BB}}(r)$ for $r_c =1.5 \sigma_{\textrm{AA}}$; magenta line: $g_{\textrm{BB}}(r)$ for $r_c =2.0 \sigma_{\textrm{AA}}$ (calculated with the cutoff at $r_c= 4.5 \sigma_{\textrm{AA}}$). \label{fig6}}
\end{figure}

Inspection of the radial distribution functions reveals that the FCS, which determines  the dynamics, is the one established by the A particles.  Figure \ref{fig6}(a) shows the three radial distribution functions for the KABLJ system.  The threshold value $r_c \approx 1.5 \sigma_{\textrm{AA}}$ for the KABLJ fluid is almost the first minimum of the radial distribution function $g_{\textrm{AA}}(r)$ of the A particles, which appears at  $r_c = 1.425\sigma_{\textrm{AA}}$. Note that this is at a distance quite different from the first minima of the two other radial distribution  function; in fact the B particles do not create a FCS of their own.

In a recent paper Rehwald, Rubner, and Heuer demonstrated that KABLJ systems of less than one hundred particles have the correct (bulk) thermodynamics and diffusivity \cite{Heuer}. Tripathy and Schweizer further very recently showed that in the activated barrier hopping theory based on naive mode-coupling theory and nonlinear Langevin equation, the short-range interactions determine the single-particle dynamics and the physical nature of the transiently arrested state (in fact, even for nonspherical particle fluids) \cite{ksch}. The present finding that it is enough to know the interactions within the FCS fits nicely into the results of these two recent works.

We end this section by briefly digressing to discuss in more detail the roles of the A and B particles. That it is the FCS around the A particles, which determines the properties of the KABLJ fluid, can be deduced from the dependence of the radial distribution function of the B particles. All three radial distribution functions depend on $r_c$, but whereas $g_{\textrm{AA}}(r)$ and $g_{\textrm{AB}}(r)$ vary only little with $r_c$ for $r_c > 1.5 \sigma_{\textrm{AA}}$ -- and in a way similar to that of the simple LJ-fluid -- the distribution of the B particles is more sensitive to $r_c$ for values larger than  $1.5 \sigma_{\textrm{AA}}$. This may seem to contradict the above conclusion, but in fact, as we shall now argue, it merely demonstrates that the A particle FCS determines the properties of the KABLJ fluid. The sensitivity of  $g_{\textrm{BB}}(r)$ to the truncation of forces  beyond  $r_c= 1.5\sigma_{\textrm{AA}}$ is clear from Fig. \ref{fig6}. Only when $r_c \geq 2 \sigma_{\textrm{AA}}$ is the correct radial distribution function obtained. Nevertheless, the msd of the B particles (Fig. \ref{KABLJdyn} (b)) and the corresponding relaxation time (blue line in Fig. \ref{KABLJdyn}(b)) agree well with the  KABLJ msd and $\tau$ already for $r_c= 1.5 \sigma_{\textrm{AA}}$. Thus the attractive forces between pairs of B particles at distances in the interval $1.5 \sigma_{\textrm{AA}} \le r\le 2.0 \sigma_{\textrm{AA}}$  changes the distribution $g_{\textrm{AA}}(r)$, but have no influence on the dynamics. A plausible explanation of this apparent contradiction is that a pair of B  particles can be bound to the same A particle within this particle's FCS and rearrange between different local minima within this FCS due to the small values $\sigma_{\textrm{AB}}=0.8 \sigma_{\textrm{AA}}$ and $\sigma_{\textrm{BB}}= 0.88 \sigma_{\textrm{AA}}$.  The local minima depend on the attraction between B particles within the A particles' FCS. Their maximum distance must be $\approx \sigma_{\textrm{AB}}+\sigma_{\textrm{AA}}=1.8 \sigma_{\textrm{AA}}$, which is slightly less than $2\sigma_{\textrm{AA}}$. Neverthelss, Figs. \ref{msd}(b) and \ref{KABLJdyn}(b) show  that this sensitivity to the attraction between pairs of B particles within the FCS of the A particle has no consequence for their dynamics.

\section{Discussion}

We have demonstrated that structure and dynamics of LJ systems in the dense fluid phase are well reproduced by introducing  a shifted-forces cutoff \cite{toxdyre} at the first minimum of the radial distribution function (a distance that approximately equals the radius of the sphere for which the density is equal to the overall density in the uniform fluid).  This means that for LJ liquids -- and presumably also for other simple liquids -- including all interactions from the particles within the FCS is all that matters for getting the correct physics. The forces from particles beyond the FCS only change the structure marginally, and their contribution to the thermodynamics can therefore be taken into account with high accuracy by simple mean-field corrections.

It is important to emphasize that this paper focused on the uniform condensed phase, far from the gas phase and the critical point. The dominant role of the forces within the FCS can only be expected to apply for the condensed phase, where a coordination shell is well established. When the density is decreased, one gradually enters a region of phase space of more gas-like behavior; here the cluster picture of the virial expansion gradually becomes more relevant and useful. For a simple LJ liquid our simulations suggest that at typical temperatures this transition takes place when $\rho \sigma^3\approx 0.6 $.

With the demonstration of the importance of the FCS for simple condensed fluids one may ask to which extent this property applies also for complex molecular liquids such as water?  Some indication that this may be the case is given in Ref. \onlinecite{Pratt}, where the free energy of a (TIP3P) water molecule is  partitioned into chemical associations with proximal inner-shell water molecules and classical electrostatic/dispersion interactions with the remaining outer-shell water molecules. The calculated free energy is in excellent agreement with  free energy per particle of (TIP3P) water, but it should be noted that of course not all interactions beyond the FCS are ignored in this approach.

The traditional discussion of the roles of attractive versus repulsive forces \cite{Widom1,bh,wca}, which was recently reinvigorated by Berthier and Tarjus \cite{ber09,ber11}, does not ask the right question. According to the presently proposed ``FCS focus'' the reason that the traditional separation into repulsive and attractive interactions fails (severely for the viscous dynamics, but actually also for the standard LJ liquid) is not {\it per se} that attractions are ignored. Rather, the dynamics and (first-order) perturbation expansion fail because at ordinary densities ignoring the attractions fails to take into account all interactions from particles within the FCS -- at extremely high densities the WCA approximation works well because here it does take into account all forces from particles within the FCS. 

More generally, the physical basis for the traditional perturbation theory \cite{bh,wca} is called into question by our findings. The idea behind perturbation theory is that repulsive and attractive forces give rise to separate, clearly identifiable contribution to the equation of state and to the free energy. Perturbation theory for the free energy is based on the physical picture that, roughly speaking, the repulsive forces decrease the entropy whereas the attractive forces decrease the energy. This picture captures the general qualitative behavior of many condensed fluids \cite{Solana}. We find, however, that in order to obtain an accurate particle distribution all forces from particles within the FCS must be included, independent of whether these forces are repulsive or attractive. As a consequence, we suggest that perturbation theory will be much more useful if the reference state as a minimum takes into account all interactions from the FCS.

The mean-field term is the first term in the Zwanzig high-temperature expansion \cite{Zwanzig}. It is exact if there is a well-defined excluded volume in the high-temperature limit. This is, however, not  the case. Our simulations at the high-temperature state point (3) demonstrate that even in this case a necessary and sufficient condition for obtaining the correct dynamics and thermodynamics is to include all forces from  the FCS -- and only these. The size of the FCS is density dependent, and no well-defined ``high-temperature limit'' exists. The Zwanzig reference system is a hypothetical reference system like the ideal gas state. As is well known, such systems are very useful for obtaining a qualitative understanding of the liquid state, as well as quantitative information, as has been demonstrated in numerous cases. Understanding the limitations of reference systems is nevertheless important for obtaining an accurate physical understanding of the physics of simple, condensed fluids.

\acknowledgments

The authors are grateful to Ben Widom for reading an early version of the manuscript and pointing out the role of the van der Waals and cell theories. We wish also to thank Trond Ingebrigtsen and Thomas Schr{\o}der for suggesting that the FCS as defined here corresponds closely the volume where the particle density equals the average density (Fig. 1). -- The centre for viscous liquid dynamics ``Glass and Time'' is sponsored by the Danish National Research Foundation (DNRF).


\begin{thebibliography}{99}
\bibitem{comment} This is so despite that fact that a SF cutoff modifies the force at all distances whereas the standard SP cutoff has correct forces for $r<r_c$. Presumably, SF works better than SP because the latter has a discontinuity in the force at $r=r_c$, which implies that the slope of the radial distribution function $g(r)$ must also have a discontinuity here (recall that the logarithm of the radial distribution function $g(r)$ gives the mean-field force potential \cite{chandler}).
\bibitem{explain} The lable LJ marks a system with $r_c=4.5 \sigma$; the data for $u$ and  $p$ are without the mean-field corrections for the integrations from $r_c=4.5$ to $\infty$.
\bibitem{kacut} The different potentials were in Ref. \onlinecite{KA} cut at $r_c = 2.5 \sigma_{\alpha, \beta}$, i.e., at varying lengths depending on the specific pair potential. The cutoff at the (WCA) minima of each pair potentials is $r_c = 1.12 \sigma_{\alpha, \beta}$. The other cutoffs are for $r_c$ in unit of $\sigma_{\textrm{AA}}$.
\end{thebibliography}
\end{document}